# Coexistence of pressure-induced structural phases in bulk black phosphorus: a combined x-ray diffraction and Raman study up to 18 GPa

B Joseph[1], S Caramazza[2], F Capitani[3], T Clarté[4], F Ripanti[2], P Lotti[1], A Lausi[1], D Di Castro[5], P Postorino[6], and P Dore[7]

[1] Elettra-Sincrotrone Trieste, S. S. 14 – km 163,5, 34149 Basovizza, Trieste, Italy
[2] Dipartimento di Fisica, Sapienza Università di Roma, P.le Aldo Moro 5, 00185 Roma, Italy
[3] Synchrotron SOLEIL, L'Orme des Merisiers, Saint-Aubin BP 48, 91192 Gif-sur-Yvette Cedex, France
[4] Département de Physique, École Normale Supérieure de Lyon, 46 Allée d'Italie, 69007 Lyon, France
[5] Dipartimento di Ingegneria Civile e Ingegneria Informatica, Università di Roma "Tor Vergata" and CNR-SPIN Via del Politecnico 1, 00133 - Roma, Italy
[6] Dipartimento di Fisica, Sapienza Università di Roma and CNR-IOM, P.le Aldo Moro 5, 00185 Roma, Italy
[7] Dipartimento di Fisica, Sapienza Università di Roma and CNR-SPIN, P.le Aldo Moro 5, 00185 Roma, Italy

**Keywords:** black phosphorus, high pressure, phase transitions, Raman spectroscopy, x-ray diffraction

**Abstract**

We report a study of the structural phase transitions induced by pressure in bulk black phosphorus by using both synchrotron x-ray diffraction for pressures up to 12.2 GPa and Raman spectroscopy up to 18.2 GPa. Very recently black phosphorus attracted large attention because of the unique properties of few-layers samples (phosphorene), but some basic questions are still open in the case of the bulk system. As concerning the presence of a Raman spectrum above 10 GPa, which should not be observed in an elemental simple cubic system, we propose a new explanation by attributing a key role to the non-hydrostatic conditions occurring in Raman experiments. Finally, a combined analysis of Raman and XRD data allowed us to obtain quantitative information on presence and extent of coexistences between different structural phases from ~5 up to ~15 GPa. This information can have an important role in theoretical studies on pressure-induced structural and electronic phase transitions in black phosphorus.

# 1. Introduction

At ambient conditions, black phosphorus (BP) is a high mobility layered semiconductor with a puckered honeycomb lattice structure, in which strong covalent bonding determines the structure of the single layers, hold together by weak van der Waals couplings. BP was the object of a number of detailed studies in the 80's, showing in particular that BP undergoes a series of pressure induced structural transitions driving the system from the orthorombic (A17) phase at ambient conditions to the rhombohedral (A7) one around 5 GPa and then into a simple cubic (sc) phase around 10 GPa [1, 2]. In the sc phase BP is metallic and a transition to a superconducting state was observed at a temperature of the order of 5-10 K. A detailed review of the BP properties investigated until 1985 is reported in Ref. [3].

In more recent years, a number of theoretical studies have been devoted to the pressure-induced structural transitions (see Ref. [4] and references therein), and the sequence of high pressure structural phases was investigated through x-ray diffraction (XRD) measurements performed up to ~350 GPa [5, 6, 7]. Studies have also been devoted to pressure effects on electronic properties [8], in particular to the pressure dependence of the superconducting transition [9, 8], and to the electronic topological transition (Lifshitz transition) occurring around 1.5 GPa in the A17 structure from a semiconducting to a semi-metal phase [10, 11].

Recently BP has attracted large attention mainly due to the demonstration of a semiconducting gap which can be tuned by varying the number of layers, from ~0.3 eV in the bulk up to ~2.0 eV in the single layer case (phosphorene). BP thus appeared to be an ideal candidate for electronic and optoelectronic applications (see the review in Ref. [14] and references therein).

Recently, the discovery of superconductivity at high pressure (300 GPa) with high transition temperature ($T_c$=200 K) in sulfur hydride [15, 16] and then in phosphorus hydride ($T_c$=100 K at 200 GPa) [17] stimulated a detailed study of the structural and superconducting phase diagram of elemental phosphorus for pressures up to 350 GPa [18]. As to structural properties, crystal structure prediction methods based on Density Functional Theory (DFT) yielded for the low-pressure regime (up to ~15 GPa) different structures with very similar enthalpy and the correct A17→A7→sc transition sequence on increasing pressure, although the pressure value at which the A7→sc transition occurs was substantially overestimated [18].

In spite of the above recent studies, some problems should be still clarified for a deeper understanding of the system properties. In particular, a coexistences between different structural phases, first observed in old XRD [19] and Raman measurements [20], was the object of recent studies [8, 21, 22], but quantitative information is still missing. Our primary aim was thus to quantitatively determine the percentage content of each phase in the coexistence regions up to ~15 GPa. Furthermore the persistence of a Raman signal above 10 GPa, as observed in old measurements [20] and confirmed by recent Raman measurements up to 24 GPa [8, 21], should be clarified since no Raman response should be observed in a cubic elemental system like BP [8, 20, 21, 22]. Our second aim was thus to try to explain the apparent contradiction between Raman and XRD results.

We performed a XRD study up to 12.2 GPa and a Raman investigation for pressures up to 18.2 GPa. A combined use of these two techniques can indeed provide clear information in the case of structural phase transitions, as shown in the case of elemental arsenic, which similarly to BP undergoes the A7→sc transition around 25 GPa [23]. Note that our XRD measurements were completed before the appearance of new XRD data [22] that fully confirmed the A7→cubic transition around 10 GPa and showed that above 10.5 GPa and up to about 30 GPa an intermediate pseudo simple cubic (p-sc) structure occurs, which is an intrinsic feature of BP as well as of other group 15 elements [24]. Our XRD and Raman measurements were performed on the same polycrystalline sample, and particular care was taken in both cases to avoid sample degradation due to prolonged air exposure [25, 26], in order to rule out a possible sample dependence in comparing results of different experiments.

## 2. Methods

Raman spectra were collected using a Horiba LabRAM HR Evolution micro-Raman spectrometer working in back-scattering geometry, installed in the High Pressure Spectroscopy Lab of the Physics Department at Sapienza University of Rome, equipped with a He-Ne laser (wavelength = 632.8 nm). For high pressure measurements performed at room temperature up to 18 GPa we employed a diamond anvil cell (DAC) and a 20x long working distance objective was used to focus the laser beam on the sample and to collect the back-scattered signal. Further technical details are reported in Ref. [27]. A thin BP crystalline platelet was scratched from a polycrystalline sample (Smart Elements, Austria) and then rapidly loaded in the sample chamber to avoid sample degradation [25, 26]. NaCl powder as hydrostatic medium and a ruby chip to determine the working pressure [28] were also loaded in the sample chamber.

XRD measurements were carried out at the Elettra Synchrotron Trieste by exploiting the characteristics of the Xpress beamline, a recently commissioned powder diffraction facility dedicated to high pressure, realized under a joint Indo-Italian collaboration. We used a large x-ray opening screw-driven plate DAC which allows a maximum pressure of ~12 GPa. Data were collected in the 5-28 2θ range by employing a 0.5 Å wavelength. Further technical details are reported in Ref. [29]. Particular care was taken in grinding a small portion of the polycrystalline sample employed for Raman, and in transferring the resulting powder in the silicone oil used as pressure transmitting medium, which was then loaded in the DAC together with ruby chips [28]. To avoid sample degradation [25, 26], the entire procedure was carried out inside a $N_2$ filled glove box.

## 3. Results

The room temperature Raman spectra are reported in Fig. 1a over the 150-500 cm$^{-1}$ frequency range, are well consistent with recent data [8, 21]. Note that a direct comparison among spectral intensities is possible since the spectra have been collected over the 0-18 GPa pressure range without changing the acquisition parameters (for clarity, the spectra in Fig. 1 are only vertically shifted). Looking at Fig. 1a, three pressure ranges can be qualitatively distinguished: *I)* up to about 6 GPa, the $A_g^1$, $B_{2g}$, and $A_g^2$ lines characteristic of the A17 orthorhombic structure [30] dominate the spectra; *II)* from ~6 to ~10 GPa the Raman intensity decreases significantly, all the Raman components remarkably broaden and a new spectral feature appears around 300 cm$^{-1}$, which progressively moves towards low frequencies on increasing the pressure; *III)* above 10 GPa, the intensity of the Raman spectrum, which continuously shifts towards low frequencies, starts to increase again, as long as a broad band centered around 250 cm$^{-1}$ dominates the Raman signal at the highest pressures.

The measured diffraction patterns are reported in Fig. 1b over the small 10.5-17.5 2θ range to highlight the most relevant modifications induced by pressure.

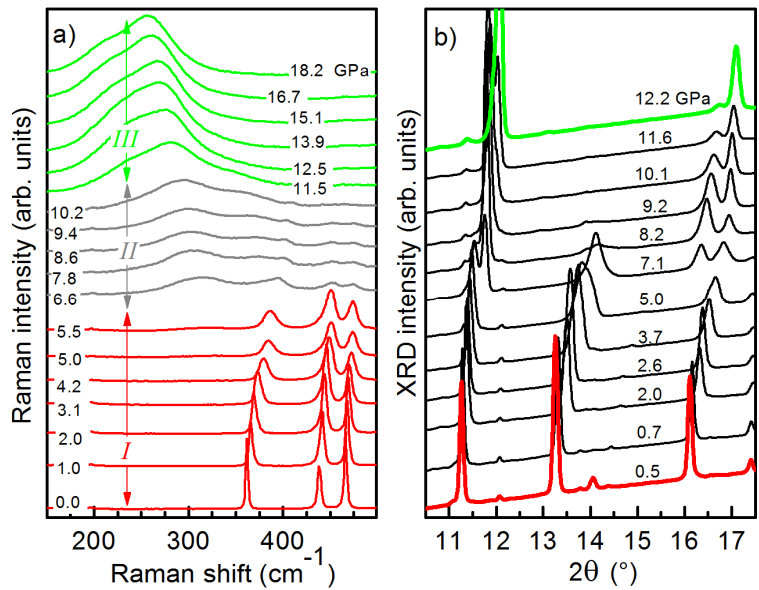

Fig.1 (Color online) (a) Pressure evolution of the Raman spectra. For clarity, the spectra are vertically shifted. Vertical arrows identify the three pressure ranges (I, II, III, see text). (b) Pressure evolution of the XRD patterns. For clarity, the patterns are vertically shifted and reported in reduced 2θ and intensity scales.

## 4. Analysis and discussion
### 4.1 Raman

The Raman spectra measured at a few representative pressures are shown in Fig. 2. Note that at P=0 (see Fig. 2a), besides the $A_g^1$, $B_{2g}$, and $A_g^2$ lines, also the low-intensity $B_{1g}$ component of the orthorombic A17 structure [20] has been observed. Considering possible effects due to the use of different lasers and polarization conditions [31], this spectrum is well consistent with previous ones measured at ambient conditions in both bulk [20, 21, 30] and few-layers systems [32].

At P=0 and up to 5.5 GPa (pressure range I), the measured spectra are well reproduced by using a standard fitting procedure by using the components characteristic of the A17 structure (see Fig. 2a). Up to about 10 GPa (range II), the broad $A_{1g}$ and $E_{1g}$ components characteristic of the rhombohedral A7 structure [20] become dominating (see Fig. 2b in the 6.6 GPa case).

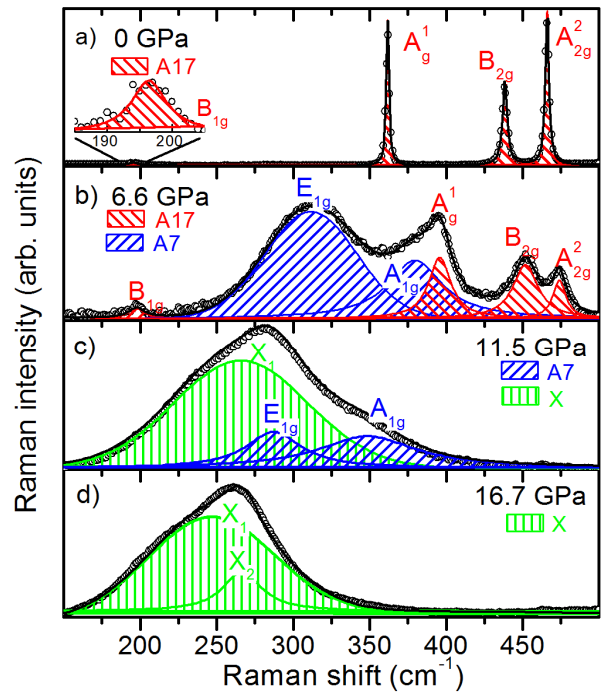

Fig.2 (Color online) (a-d) Raman spectra measured at representative pressures and best fit profiles. Phonon components ascribed to the A17, A7 and X structures (see text) are evidenced.

Above 10 GPa (range III), the expected structural transition to the sc phase should produce the complete disappearance of the Raman spectrum, whereas the overall Raman intensity is significantly increased (see Fig. 1a). Since the obtained results (see Fig. 2c at 11.5 GPa and Fig. 2d at 16.7 GPa) cannot be attributed to a sc structure nor to a persistence of the A7 structure above 10 GPa, we tentatively introduced a new unidentified structure, here and in the following named as X phase. At 11.5 GPa a $X_1$ component is much more intense than those due to a persisting A7 structure (see Fig. 2c), while above 15 GPa the structure appears to be nearly stabilized with only the presence of only two components ($X_1$ and $X_2$, see Fig. 2d in the 16.7 GPa case). Note that above 10 GPa the pressure dependence of the phonon frequencies we determined is quite similar to that recently reported and discussed (see Fig. 5 in Ref. [21]), where the two peaks above 15 GPa are attributed to the sc phase, while we attribute them to the X phase.

**4.2 XRD**

The XRD patterns are reported at a few representative pressures in the extended 8-28 2θ range in Figs. 3a-c, in a limited 11.2-12.3 2θ range around the most intense lines in Figs. 3d-g.

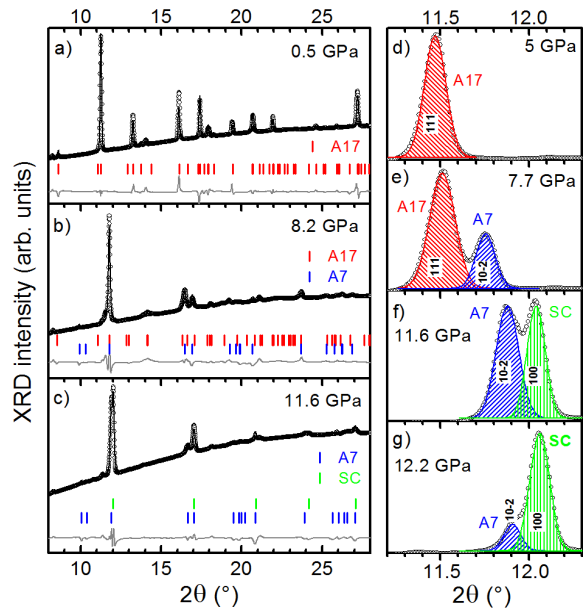

Fig.3 (Color online) (a-c) XRD patterns at representative pressures and best fit profiles in the extended 8-28 2θ range. (d-g) XRD intensity at representative pressures in the limited 11.2-12.3 2θ range, where components due to different structures are evidenced.

Structural refinements of the diffraction data were performed using the General Structure Analysis Software (GSAS), a well-known software package for crystallography data analysis [33]. For pressures up to 5 GPa only the A17 structure was required for describing the diffraction patterns, as shown in Fig. 3a (0.5 GPa) in the extended range and in Fig. 3d (5 GPa) in the limited range. Above 5 GPa, good fits were obtained only by using a two-phase refinement involving the A17 and A7 structures, as shown in the 8.2 GPa (Fig. 3b) and 7.7 GPa (Fig. 3e) cases. Finally, above 10 GPa a good description of the diffraction data demanded a two-phase refinement involving both A7 and sc structures, as shown at 11.6 GPa (Fig. 3c and Fig. 3f) and at 12.2 GPa (Fig. 3g). In summary, the coexistence between A17 and A7 structures from ~5 up to ~10 GPa obtained from Raman analysis is fully confirmed by the present XRD study. Moreover, above ~10 GPa the presence of the sc structure, coexisting with the A7 one, is clearly identified.

**4.3 Discussion**

First of all we want to discuss the apparent contradiction between XRD and Raman results, since the simple cubic phase, observed in previous [1, 2, 8, 22] and present XRD data, is not compatible with the intense spectrum well evident above 10 GPa in present and previous Raman data [8, 20, 21].

In previous works, these Raman components have been attributed to different disorder-induced effects [8, 21]. Also in the case of elemental arsenic, the Raman components observed by Beister *et al.* at pressures above the A7→sc transition [23] were attributed to disorder-induced scattering. However, the authors clearly note that these components are weak and decrease with increasing pressure [23]. On the contrary, the intensity of the Raman spectrum we observe in BP increases on increasing pressure in the sc phase (see Sect.3 and Fig.1a). Therefore, in our opinion the Raman spectrum we observe in the sc phase cannot be attributed to disorder-induced effects.

In principle, also the p-sc structure occurring in BP above 10.5 GPa [22, 24] could be responsible of the appearance of Raman components. In ref. [22] the p-sc phase is identified through the occurrence of two weak Bragg peaks, apart from the four prominent peaks of the sc phase. Even if we did not observe these two weak peaks due to low signal-to-noise ratio, we cannot rule out the possibility that the high pressure phase is p-sc rather than sc. However, according to ref [22], the distortion of the ideal sc structure (in which no Raman spectrum can be observed) is very small in the p-sc case (see Fig.4 of ref [22]) and decreases on increasing pressure until the ideal cubic phase is reached around 30 GPa [22, 24]. On the contrary, our data (in agreement with recent literature results [8, 21]) show that the intense Raman spectrum we observe above 10 GPa is strongly pressure dependent and increases on increasing pressure (see fig. 2a). On this basis, in our opinion it is not reasonable to attribute the high pressure Raman spectrum only to the presence of the p-sc phase.

To explain the apparent contradiction between XRD and Raman results, and in absence of other plausible scenarios, we propose a different hypothesis, based on the experimental conditions occurring in the two different experiments. In the XRD case, the DAC cell is loaded with silicone oil containing the powder sample. This procedure can guarantee a hydrostatic condition, i.e. an isotropic pressure on the sample grains. On the contrary, as usual in micro-Raman measurements to maximize the back-scattered Raman signal, a small crystalline platelet is pressed between diamond window and hydrostatic medium. Consequently, a hydrostatic condition is presumably not achieved. Since DFT computations [18], based on the assumption of a hydrostatic pressure, show that different structures are energetically competitive up to ~15 GPa, it is reasonable to hypothesize that the sequence of structures in XRD experiments (hydrostatic conditions) can be different from that occurring in Raman experiments (non-hydrostatic conditions).

In order to determine the quantitative presence of each structure in the coexistence regions, which was the main aim of the present work, we considered both Raman and XRD results. In the case of Raman, the fitting procedure performed at each pressure (see Figs. 2a-d) allowed the identification of the $k$-components associated to a given $i$-structure ($i$ =A17, A7, X) and thus the evaluation of the corresponding integrated area (spectral weight $sw_k$). By summing the $sw_k$ of the $i$-structure and dividing by the integrated area of the entire spectrum, we determined at each pressure the percentage spectral weight of the $i$-structure (structure $PSW_i$, see Fig. 4a). $PSW_i$ is proportional to the amount of the $i$-structure and in principle it might provide the percentage content of the $i$-structure. In the case of XRD data, the two-phase structural refinement performed at each pressure by using the GSAS package [33] directly provided the percentage content of the i-structure (structure PCi, i =A17, A7, sc, see Fig. 4b). Note that the PC value thus obtained above 10 GPa is very similar to the relative weight of the two Bragg peaks around 2θ ~ 11.8.

A direct comparison between the structure $PSW_i$ from Raman (see Fig. 4a) and the structure $PC_i$ from XRD (see Fig. 4b) is possible by assuming that the Raman cross sections for different structures are coincident. This assumption is not acceptable, since Raman cross sections depend on different factors, such as the Raman tensors. However, we note that the pressure dependences of $PSW_i$ and $PC_i$ are consistent and coherently show how the A17 structure is progressively substituted by the A7 above ~5 GPa. Above ~10 GPa, the increase of the $PC_{sc}$ from XRD quite resembles the increase of $PSW_X$ from Raman (see Fig. 4). Consequently, we can argue that the pressure increase above 10 GPa drives the sample grains to the sc structure (XRD case), the crystalline platelet to the unidentified X structure (Raman case).

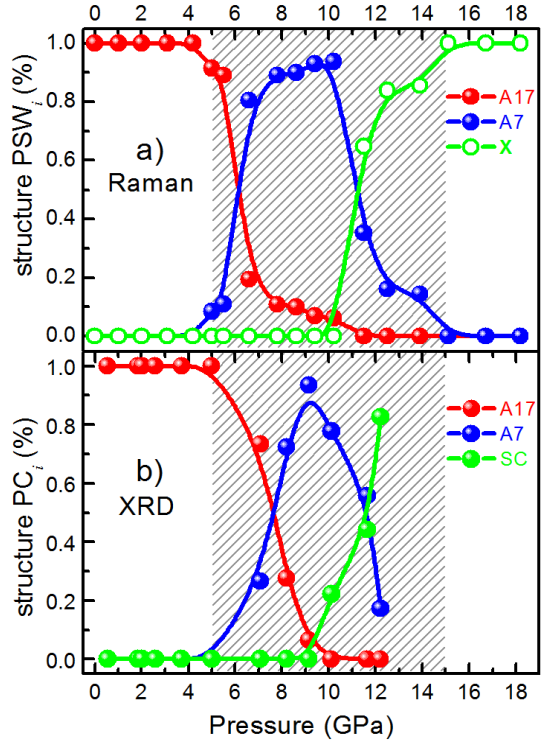

Fig.4 (Color online) (a) Pressure dependence of the percentage spectral weight of the i-structure PSWi (i = A17, A7, X) from Raman analysis (see text). (b) Pressure dependence of the percentage content of the i-structure PCi (i = A17, A7, sc) from XRD analysis (see text). The dashed area marks the pressure range where coexistence between different phases occurs.

## 5. Conclusions

In the present paper we report a study of the structural phase transitions induced by pressure in bulk black phosphorus by using both x-ray diffraction for pressures up to 12.2 GPa and Raman spectroscopy up to 18.2 GPa. The powder for XRD and the crystalline platelet for Raman were obtained from the same polycrystalline sample. We first confirm the presence of Raman components above 10 GPa, that are unexpected since no Raman components should be observed in the simple cubic phase evidenced by present and previous XRD data. In our opinion this apparent contradiction may be explained by attributing a key role to the non-hydrostatic conditions occurring in Raman experiments, which can prevent the transition from the A7 to the sc structure in the crystalline sample, which above 15 GPa appears to be finally stabilized in a structure characterized by an intense and broad two-component Raman spectrum centered around 250 cm$^{-1}$. After the present work was completed, we became aware of the recent paper by Sasaky et al. [34] which reports Raman spectra of few-layers BP films under high pressures, showing broad and intense Raman components below 300 cm$^{-1}$ for pressures above 10 GPa. This work supports our hypothesis, since the authors in discussing their results suggest a key role of non-hydrostaticity of pressure.

In conclusion, a combined analysis of Raman and XRD data allowed us to establish presence and extent of coexistence between different structural phases in a wide pressure range (from ~5 up to ~15 GPa, see Fig. 4). These findings, and in particular the evidence of a non-complete conversion to the simple cubic phase even at 12.2 GPa, can have an important role also in theoretical studies on pressure-induced electronic phase transitions on bulk BP.